\documentclass[conference]{IEEEtran}
\usepackage{cite}
\usepackage{amsmath,amssymb,amsfonts}
\usepackage{algorithm}
\usepackage{algorithmicx}
\usepackage{algpseudocode}
\usepackage{graphicx}
\usepackage[export]{adjustbox}
\usepackage{textcomp}
\usepackage{xcolor}
\usepackage{subfigure}
\usepackage{csquotes}

\let\svthefootnote\thefootnote

\def\BibTeX{{\rm B\kern-.05em{\sc i\kern-.025em b}\kern-.08em
    T\kern-.1667em\lower.7ex\hbox{E}\kern-.125emX}}
\begin{document}

\title{Value of Information and Timing-aware Scheduling for Federated Learning}

\author{\IEEEauthorblockN{Muhammad Azeem Khan\IEEEauthorrefmark{1}, Howard H. Yang\IEEEauthorrefmark{2}, Zihan Chen\IEEEauthorrefmark{3}, Antonio Iera\IEEEauthorrefmark{1}, and Nikolaos Pappas\IEEEauthorrefmark{4}}	
	\IEEEauthorblockA{\IEEEauthorrefmark{1}DIMES Department, University of Calabria, Arcavacata di Rende (CS), Italy \\ \IEEEauthorrefmark{2}ZJU-UIUC Institute, Zhejiang University, China\\ \IEEEauthorrefmark{3}ISTD Pillar, Singapore University of Technology and Design, Singapore \\\IEEEauthorrefmark{4}Department of Computer and Information Science Link\"oping University, Sweden\\
		Email: khnmmm95d04z236f@studenti.unical.it, haoyang@intl.zju.edu.cn, \\zihan\_chen@sutd.edu.sg, antonio.iera@dimes.unical.it, nikolaos.pappas@liu.se
}}

\maketitle

\begin{abstract}
Data possesses significant value as it fuels advancements in AI. However, protecting the privacy of the data generated by end-user devices has become crucial. Federated Learning (FL) offers a solution by preserving data privacy during training. FL brings the model directly to User Equipments (UEs) for local training by an access point (AP). The AP periodically aggregates trained parameters from UEs, enhancing the model and sending it back to them. However, due to communication constraints, only a subset of UEs can update parameters during each global aggregation. Consequently, developing innovative scheduling algorithms is vital to enable complete FL implementation and enhance FL convergence. In this paper, we present a scheduling policy combining Age of Update (AoU) concepts and data Shapley metrics. This policy considers the freshness and value of received parameter updates from individual data sources and real-time channel conditions to enhance FL's operational efficiency. The proposed algorithm is simple, and its effectiveness is demonstrated through simulations.
\end{abstract}

\begin{IEEEkeywords}
Age of Update, Data Shapely, Federated Learning, Scheduling.
\end{IEEEkeywords}

\section{Introduction}
\let\thefootnote\relax\footnotetext{This work was performed when M. A. Khan was visiting Linköping University as an Erasmus student. The work of N. Pappas has been supported in part by the Swedish Research Council (VR), ELLIIT, Zenith, and the European Union (ETHER, 101096526).}
\let\thefootnote\svthefootnote

The rise in processing capabilities of end-user devices and the escalating focus on safeguarding the privacy of data, which serves as the driving force behind AI development, have triggered a significant transformation in the design and execution of machine learning models. Previously, complex computations were primarily conducted in centralized cloud environments. However, with the advancements in device capabilities and the desire to preserve data privacy, there is a transition towards performing these computations at the edges of networks, closer to where the data is generated and used. This convergence of AI and edge computing has given rise to a new approach known as Federated Learning (FL), where models are trained collaboratively across multiple devices while keeping the data decentralized and private \cite{konevcny2016federated, mcmahan2017communication, lin2017deep, zhao2020federated, wang2019adaptive, liu2021adaptive, letaief2019roadmap}. FL enables collaborative training of statistical models between a central Access Point (AP) and distributed User Equipments (UEs). This training process occurs directly on UEs using locally stored datasets while safeguarding data privacy. 

In contrast to traditional machine learning, where data is aggregated in a central location for training, FL brings the training process to UEs, with only model/gradient parameters exchanged between the AP and UEs to improve the global model \cite{mcmahan2017communication}. This iterative process continues until the global model converges. The potential for improved privacy and decreased communication overhead makes FL suitable in the context of next-generation mobile networks \cite{chen2020joint}. In this procedure, the wireless channels vary over time, leading to only a portion of UEs being able to form reliable links with the AP during each round of training. Additionally, due to the constraints of limited spectral resources, the AP can only select a subset among the reliable UEs for participation in each communication round of the FL process.

Recognizing the importance of these factors, extensive research has been undertaken, resulting in various scheduling protocols for FL. These protocols range from reducing transmission delay \cite{nishio2019client} and maximizing spectrum usage \cite{yang2020federated} to cleverly alternating between choosing UEs with different channel conditions \cite{zhu2019broadband}. Some studies have also considered update staleness, a crucial factor impacting FL's convergence rate compared to other scheduling methods \cite{yang2020age, 9484497, wang2022age,liu2021age}. Although these studies have demonstrated favorable results, they tend to disregard a crucial aspect: the value of information contributed by individual data sources engaged in each round during FL training. As emphasized by \cite{ghorbani2019data, tang2021data, 9006179}, prioritizing high-value data sources substantially enhances the convergence rate of decentralized machine learning models such as FL.
Consequently, it becomes reasonable to anticipate the development of a scheduling algorithm that considers both the value and freshness of information in addition to communication quality. This holistic approach is expected to accelerate the convergence of FL within a mobile edge network.

Our results indicate that in the presence of reliable communication channels, the convergence rate of FL can be improved through two approaches:
\begin{itemize}
    \item Expanding the communication bandwidth to select more UEs in each round.
    \item Deploying a scheduling strategy aimed at minimizing parameter staleness, such as Age of Information-based Scheduling, or prioritizing the selection of more valuable UEs based on information value.
\end{itemize}
To attain this goal, we introduce a novel strategy that integrates the concept of non-linear Age of Update for measuring update staleness and employs Data Shapley value for equitable evaluation of each UE's contribution. Utilizing these concepts, we design a scheduling policy that empowers the access point (AP) to efficiently collect valuable and timely updates from UEs for FL training. Considering scenarios where communication channels demonstrate significant unreliability, the approaches mentioned above may not result in considerable benefits for enhancing the efficiency of FL training.

We outline the model framework for our proposed solution in Section II. Section III provides an overview of the Age of Update (AoU) and Data Shapley metrics employed in our scheduling policies. In Sections IV and V we discuss implementation of the AoU and Data Shapley metrics and detail the development of a scheduling policy that combines the concepts of AoU and Data Shapley for our proposed solutions, respectively. In Section VI, we examine the performance of our proposed solution through a comparative analysis of simulation results against alternative approaches. Our contributions and potential options for further research are summarized in Section VII.

\begin{figure}[t]
    \centerline{\includegraphics[width=1\linewidth]{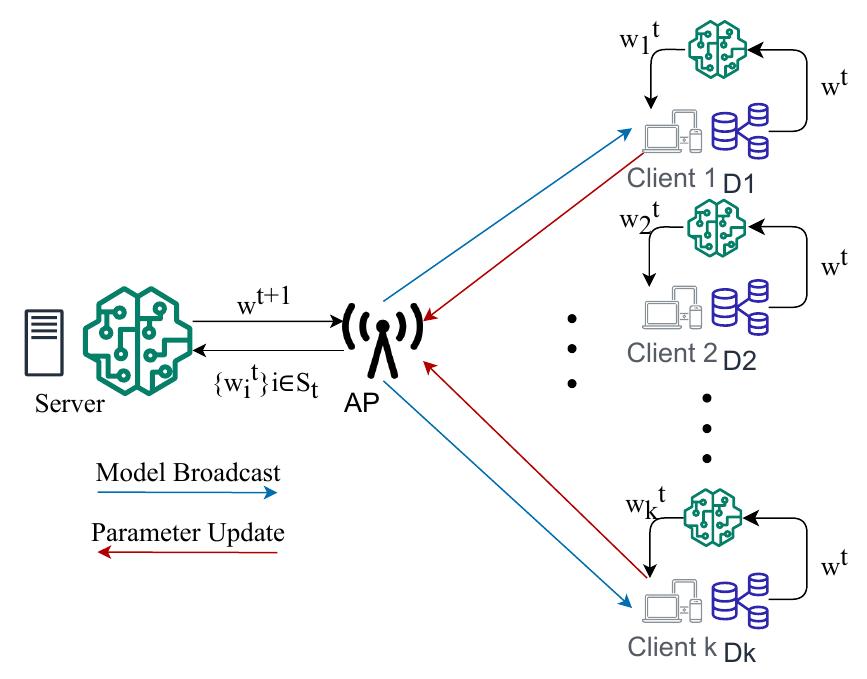}}
    \caption{An illustration of the federated learning process: collaboration of edge devices for model training.}
    \label{fig:Figure1}
\end{figure}

\section{Model Framework}
Let's examine an FL setup comprising an access point (AP) and $K$ clients, with $K$ being a substantial count, all clients are assumed to have a single antenna and possess the capability to conduct computations locally as illustrated in Fig.\ref{fig:Figure1}. Throughout this paper, we will interchangeably use the terms \enquote{User Equipment} (UE) and \enquote{client}.
Each client $k$ possesses its distinct local data set denoted as $D_k$, where; $D_k = \{(x_i, y_i)\}_{i=1}^{K}$. We presume that these individual datasets are statistically uncorrelated among the clients. Within this network setup, the client $k$ establishes communication with the AP via a shared spectrum, partitioned into N subchannels of uniform length. It's important to note that $N \ll K$. 

We adopt a Rayleigh fading propagation model, where the channels between any pair of antennas are assumed to follow the Rayleigh distribution and exhibit quasi-static behavior i.e. the channel remains constant during a transmission in a communication round but can vary independently from one round to another. The objective of the AP is to develop a statistical model that encompasses all client datasets while upholding their privacy. In the study conducted by \cite{9652996}, the AP is tasked with optimizing the model parameter $w \in \mathbb{R}^d$. This optimization aims to minimize the following loss function, even in the absence of direct access to the datasets $D = \{D_k\}_{k=1}^{K}$: 
\begin{equation}
    \min_{w \in \mathbb{R}^d}\{f(w)= \frac{1}{n} \sum_{i=1}^{n} l(w; x_i, y_i)
    =\sum_{k=1}^{K} p_kf_k(w)\} 
    \label{eq:objective function}
\end{equation}
where $n = \sum_{k=1}^{K}n_k$, and $l(.)$ is the loss function defined under some criteria, $p_k = {n_k}/n$ and $f_k(w)$ denotes the local empirical loss function of client $k$, given by
\begin{equation}
    f_k(w) = \frac{1}{n_k} \sum_{j=1}^{n_k} l(w;x_j,y_j).
\end{equation}
In FL, where the AP lacks direct access to clients' datasets, collaborative training occurs. After a substantial number of training iterations and the exchange of updates, referred to as \enquote{communication rounds} and \enquote{time slots}, between the AP and client, the objective function Eq.\ref{eq:objective function}, typically converges towards the global optimum. A \enquote{communication round} is a full training cycle involving interactions, synchronization, and update aggregation. Conversely, \enquote{time slots} are smaller, granular steps within rounds, fine-tuning local computations to enhance global model convergence. The implementation of communication rounds and time slots varies based on the specific FL algorithm and framework.
However, due to limitations within the wireless medium, the AP is constrained to select a subset of clients for parameter updates during each round. The selection of clients significantly impacts the pace of FL convergence rate [19]. As a result, the demand for effective scheduling algorithms in this scenario becomes a critical imperative.

\section{Metrics for Scheduling Policies}
The scheduling of UEs for parametric update in each communication round is implemented by taking into account two selection metrics: the \enquote{Data Shapley value} and the \enquote{Age of Update}. In this section we will discuss the concepts and significance of these selection metrics and elaborate them in detail.

\subsection{Age of Update}\label{AoU}
To assess the degree of update staleness, we make use of the concept of information freshness, often referred to as the Age of Information (AoI) \cite{Kosta2017age,8930830,9380899, pappas2023age}. AoI represents the time elapsed since the last successfully received update packet was generated at the source and received by the monitor. Building upon the AoI concept, \cite{yang2020age} introduced a metric known as \textit{Age of Update} (AoU). For a given UE $k$, its AoU is defined as follows: 
\begin{equation}
    T_k[t+1] = (T_k[t]+1)(1-S_k[t]); \quad S_k[t] \in \{0,1\},
    \label{eq:Original AoU}
\end{equation}

where $T_k[0] = 1$, and $S_k[t]$ takes value 1 if UE k is selected by the AP for update during communication round $t$, and takes value 0 otherwise. 

\subsection{Data Shapley}\label{DS}
Data Shapley, rooted in cooperative game theory, offers a natural solution to the challenge of fairly distributing value among individual data samples, and consequently the data sources in machine learning models. This becomes particularly essential when dealing with data from diverse sources. It assigns values based on how a data point affects the model's performance when added or removed. Particularly useful in FL, Data Shapley ensures equitable compensation for UEs' contributions, aiding in UE selection, model efficiency, and privacy concerns.

The essence of machine learning rests upon three key elements, the training dataset denoted as D = ${(x_i, y_i)}_n$, where $n$ signifies the number of distinct data sources. Each source is represented by $(x_i, y_i)$, denoting the i'th data source. 
The second key ingredient is the learning algorithm, denoted as $A$, that takes the input of the training data set $D$ and outputs a predictor.
Lastly, the evaluation of the acquired predictor plays a pivotal role. This assessment is quantified through a performance score labeled as $V$, which takes any predictor as input and returns a corresponding score. 

In \cite{ghorbani2019data}, this score is referred to as $V(S, A)$, or simply $V(S)$, to symbolize the performance rating of the predictor trained using the learning algorithm $A$ on the training data subset $S$ where $S \subseteq D$. The objective is to ascertain the value of each data source $(x_i, y_i) \in D$, referred to as $\phi_i(D, A, V)$, or simply $\phi_i$ to simplify notation. As outlined in \cite{ghorbani2019data}, in order to achieve equitable assessment of a data source, any approach must satisfy the following criteria:
\begin{itemize}
    \item If the inclusion of the $i$-th data source in any subset of the training data sources does not lead to any change in performance, it should be assigned a value of zero. To elaborate, if for all $S \subseteq D - {i}$, it holds that $V(S) = V(S \cup {i})$, then $\phi_i = 0$.
    \item If, for data sources $i$ and $j$, and any subset $S \subseteq D - {i, j}$, we find that $V(S \cup {i}) = V(S \cup {j})$, then $\phi_i = \phi_j$. In simpler terms, if data sources $i$ and $j$ provide identical contributions when added to any subset of our training data sources, they should be assigned the same value due to symmetry.
    \item When the collective performance score can be expressed as the sum of individual performance scores, the overall value attributed to a source should also be the sum of its values for each score: $\phi_i(V + W) = \phi_i(V) + \phi_i(W)$, where $V$ and $W$ represent performance scores. If source $i$ contributes values $\phi_i(V_1)$ and $\phi_i(V_2)$ to predict test points 1 and 2, respectively, we anticipate that the value of source $i$ in predicting both test points, denoted by $V = V_1 + V_2$, would be $\phi_i(V_1) + \phi_i(V_2)$.
\end{itemize}

\section{Implementation of AoU and Data Shapley }\label{Implementation of AoU and DS}
Drawing on the concept of Age of update as introduced in \cite{yang2020age}, Eq.\ref{eq:Original AoU}; this investigation incorporates a non-linear Age of Update by substituting the fixed value of 1 with $x^2$ in Eq.\ref{eq:Original AoU}. In this context, $x$ can take on various functional forms depending on the specific application, where, in our study, we assume $x$ as a function of time slots. Non-linear aging functions were introduced in \cite{Kosta2017b, Kosta2020}.
Consequently, the AoU for a typical client $k$ exhibits exponential growth. This is mathematically represented by Eq.\ref{eq:Non linear AoU}:
\begin{equation}
T_k[t+1] = (T_k[t] + x^2)(1-S_k[t]); \quad S_k[t] \in {0,1}.
\label{eq:Non linear AoU}
\end{equation}

In practical terms, this means that initiating from an AoU of 1, the AoU for client $k$ at time $t+1$ is updated as follows:
\begin{itemize}
    \item If $S_k[t] = 0$, signifying that the client $k$ didn't share update with the AP, their AoU increases by $x^2$.
    \item If $S_k[t] = 1$, indicating that the client $k$ shared updates with AP, so their AoU goes back to its initial value
\end{itemize}

This enhancement introduces non-linear growth in the age, differing from the original Eq.\ref{eq:Original AoU}, where the increment was a constant value of 1. The extent of the age increment now hinges on the magnitude of $x$. A larger $x$ corresponds to a more substantial increase due to non-updates, while a smaller $x$ results in a comparatively smaller age increment.

Additionally, we execute the three criteria introduced in the Data Shapley framework in \ref{DS} to distribute values impartially among each data source or UE based on a performance score $V$. Within the scope of this research, the score $V$ is defined as the accuracy variation between the present and preceding communication rounds during the course of FL training. These criteria are executed in accordance with the subsequent principles:
\begin{itemize}
    \item The initial values $\phi_0$ assigned to UEs are randomly generated within the range of 0 to 1.
    \item The algorithm we propose consistently monitors the performance score $V$ throughout the FL process, documenting the previous values of individual UE $k$ in a dedicated dictionary for every communication round $r:$    
    \[
    \phi_{k,r} = \{(s, \phi_{k,r}) \, | \, 1 \leq s \leq \text{r}\}.
    \]
    \item For UEs $k$ partaking in round $r$, if $V$ surpasses a predetermined threshold $\gamma$, their earlier value is incremented by 1:
    \[
    \phi_{k,r} = \begin{cases}
    \phi_{k,r-1} + 1, & \text{if } V > \gamma \\
    \phi_{k,r-1}, & \text{otherwise.}
    \end{cases}
    \]
    In contrast, if either a UE doesn't participate or $V$ remains below the threshold, the UE retains its previous value for that round:
    \[
    \phi_{k,r} = \phi_{k,r-1}.
    \]
    \item At the conclusion of each communication round $r$, the latest value assigned to each UE is calculated as the mean of its previous values:
    \[
    \phi_{k,r} = \frac{1}{r} \sum_{s=1}^{r} \phi_{k,s}.
    \]
\end{itemize}

By adhering to these stipulated criteria, we leverage the Data Shapley methodology to assign impartial values to UEs, thereby fostering enhancements in the convergence rate of Federated Learning.

\section{Designing a Scheduling Policy}\label{Designing a Scheduling Policy}
In this study, our focus lies in the communication dynamics between the AP and UEs within a resource-constrained spectrum, characterized by varying channel gains over time. Our primary objective is to design a scheduling policy that optimizes the AP's collection of updates, emphasizing both the freshness and value of the information being gathered. We build upon the foundational assumptions laid out in prior research, where UEs establish connections with the AP with a probability $p$ during global aggregations, and these connections evolve independently across communication rounds. Additionally, the server is equipped with prior knowledge about the reliability of client connections at the initiation of each global aggregation $t$.

Consequently, this leads to the central task of the server: selecting a subset $S_t$ from the client pool to participate in FL training; where $|S_t| \leq N \ll K$ , and $S[t] = \{ S_1[t],S_2[t],...,S_k[t]\}$. This selection process is achieved by minimizing specific AoU related functions and communication parameters, all while incorporating the innovative concept of Data Shapley value. The subsequent process of selecting UEs for updates post-global aggregation in round $t$ is outlined as follows:

\begin{subequations}
\begin{align}
    \min_{\substack{S_k[t], P}} \left\{ \sum_{k=1}^{K} (T_k[t] + x^2)(1-S_k[t]) \right\} \\
    \label{eq:subeq-b}
    \text{s.t.} \quad \left(\log(1 + G_{k,n}P_{k,n}) \right) S_k[t] \geq 0 \\
    \label{eq:subeq-c}
    P_{k,n} \leq P_{\text{TX}} \\
    \label{eq:subeq-d}
    {S_k[t] \in \{0,1\} \quad \forall k \in \{1,\ldots,K\}}.
\end{align}
\end{subequations}
In this context $G_k,n$ represents the gain attainable by client $k$ on the $n-th$ sub-channel, and $P_k,n$ is the corresponding injected power. The inequality (\ref{eq:subeq-b}) guarantees a positive or zero signal-to-noise ratio (SNR) on $n-th$ sub-channel for UE $k$, ensuring reliable communication. A low or negative SNR could indicate an unreliable channel, potentially hampering effective update transmission. The system also imposes limitations on the maximum transmit power, outlined in constraint (\ref{eq:subeq-c}), which helps in regulating power usage and ensuring efficient communication within the network. Once the set of reliable UEs is identified based on the aforementioned criteria, our proposed algorithm schedules the clients as follows: 
\begin{enumerate}
    \item[A)] If the number of UE with reliable channels is less than $N$, all of them will be chosen for parameter update.
    \item[B)] Otherwise arrange all the reliable UEs in descending order within a UE list, influenced by both AoU and Data Shapley values of the UEs. The algorithm then selects the first $N$ UEs (where $N$ aligns with the number of available sub-channels) for each communication round. This arrangement can be determined using one of the following distinct criteria, each of which was evaluated independently:
\begin{enumerate}
    \item[1.] Using AoU OR Data Shapley Value: If a UE's $k$ AoU surpasses the threshold or its Data Shapley value exceeds the Shapley value of the current highest value in the list, or both conditions are met, the UE $k$ is positioned at the beginning of the list; otherwise, it is placed at the end.
    \item[2.] Using AoU AND Data Shapley Value: If the AoU of a UE $k$ exceeds the threshold value and also its Data Shapley value is greater than the Shapley value of the current highest value in the list, only then the UE $k$ is placed at the beginning of the list; otherwise, it is placed at the end of the list.
\end{enumerate}
\end{enumerate}
By considering these criteria, we utilize jointly the AoU and Data Shapley values to prioritize UE selection for each communication round. This ensures that always an optimized and efficient subset of UEs is chosen.
We would like to emphasize that among the aforementioned criteria, the \enquote{AoU OR Data Shapley value} criterion outperforms the others. As we will show in the next section, Criterion 1 significantly accelerates the convergence rate of Federated Learning.

\begin{figure*}
\centering
\subfigure[]{\includegraphics[width=0.85\columnwidth]{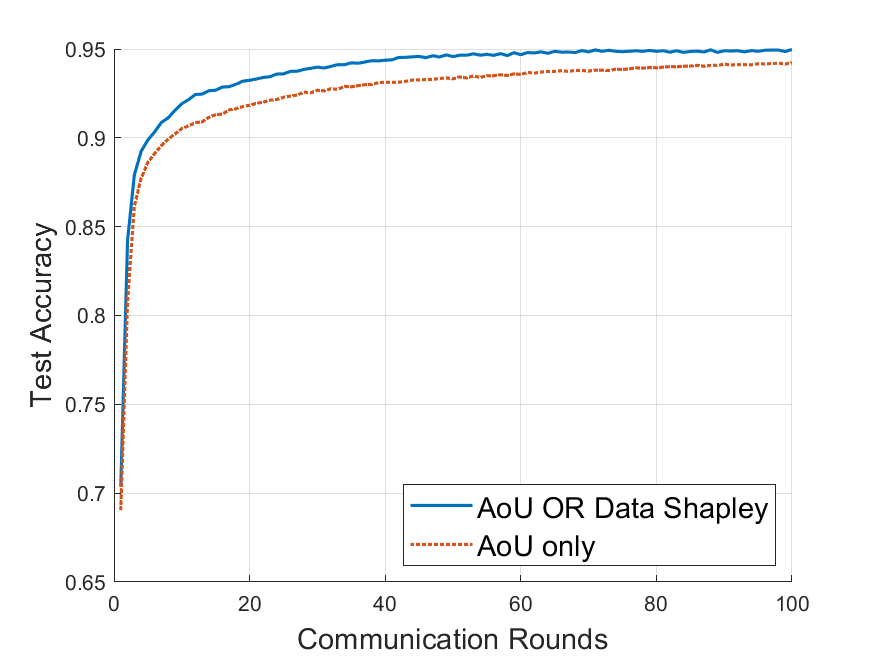}}
\hfil
\subfigure[]{\includegraphics[width=0.85\columnwidth]{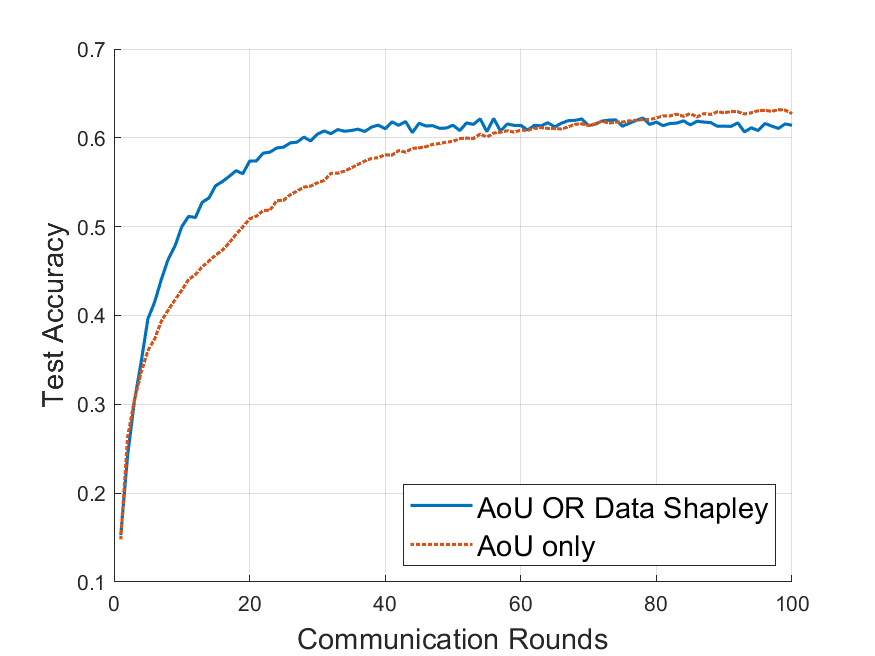}}
\caption{Comparison of FL training convergence rates: AoU OR Data Shapley vs AoU-only scheduling policies. Both scenarios with p = 0.8 and N = 30 communication channels. (a) MLP MNIST dataset with i.i.d. client assignment. (b) CNN CIFAR-10 dataset.}
\label{fig:figure2}
\end{figure*}

\begin{figure*}
\centering
\subfigure[]{\includegraphics[width=0.85\columnwidth]{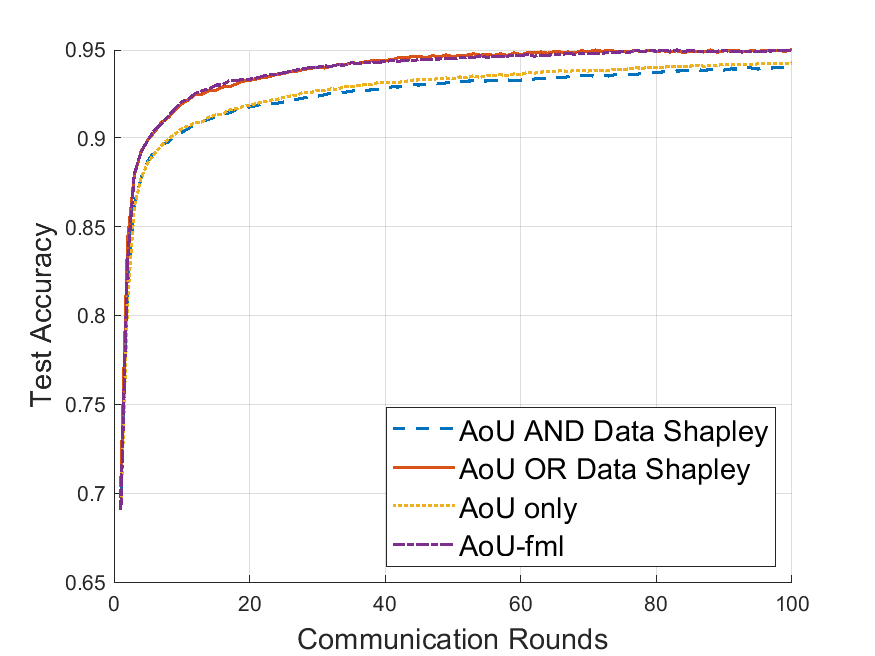}}
\hfil
\subfigure[]{\includegraphics[width=0.85\columnwidth]{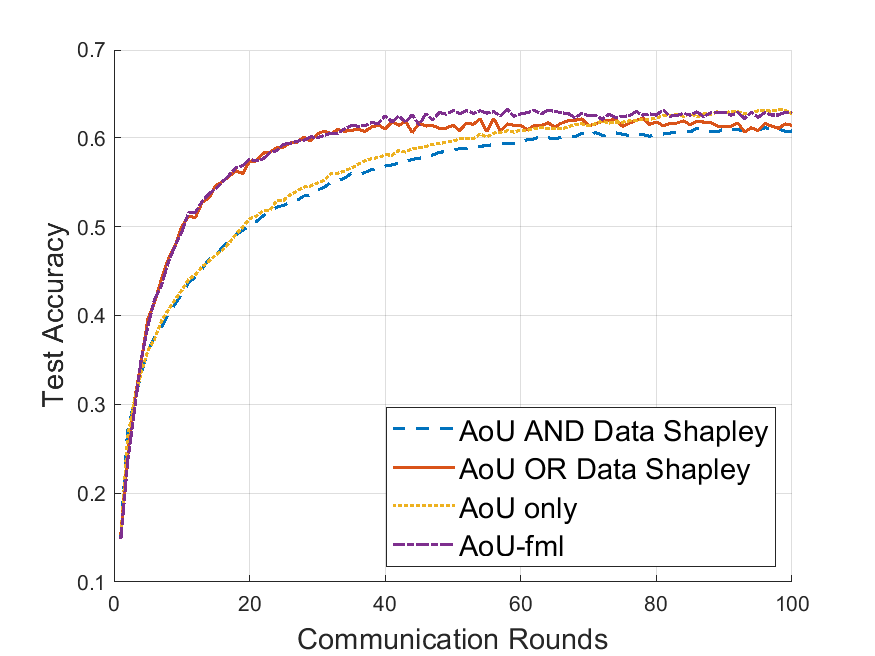}}
\caption{Comparison of FL training convergence rates under different scheduling policies. Both scenarios with p = 0.8 and N = 30 communication channels. (a) MLP MNIST dataset with i.i.d. client assignment. (b) CNN CIFAR-10 dataset.}
\label{fig:figure3}
\end{figure*}

\section{Simulation Results}
In this section, we present early simulation outcomes that validate our developed analysis. We specifically explore the efficacy of FL training using two distinct configurations of machine learning models. The initial experiment involves training an MLP with the MNIST dataset. This MLP comprises two hidden layers with 64 units each, utilizing the ReLU activation function.

The second experiment focuses on training a Convolutional Neural Network CNN on the CIFAR-10 dataset. The CNN architecture encompasses two convolutional layers with max pooling, two fully connected layers, and a softmax output layer.

We employ the methodology outlined in \cite{mcmahan2017communication} to split the entire training dataset into 100 non-overlapping segments distributed among $K = 100$ UEs. In our CIFAR-10 dataset experiments, we exclusively use independent and identically distributed i.i.d. distributions due to the absence of a natural data user partition. However, for the MNIST dataset, we explore two data partitioning approaches: i.i.d. and non-i.i.d. For i.i.d. local data partitioning, we randomly allocate the dataset uniformly across UEs. In contrast, for non-i.i.d. data partitioning, we employ a labeling-based scheme, dividing the data into $200$ shards and assigning each UE two shards. This allows us to assess the performance of our algorithms under highly non-i.i.d. data conditions. Furthermore, we consider wireless channels to be reliable with a probability of $p = 0.8$ and less reliable with a probability of $p = 0.1$. All the experiments are performed using the PyTorch framework.

Fig.\ref{fig:figure2} reveals that in scenarios of reliable communication channels (with a probability of $0.8$), our novel scheduling strategy, which combines AoU and Data Shapley value, outperforms the traditional approach solely considering AoU for UE scheduling. This refined scheme accelerates the convergence rate within FL.

In Fig.\ref{fig:figure3}, a comprehensive comparison of different UE scheduling policies is presented. This includes scenarios where only AoU is considered, a federated momentum learning approach (AoU fml) is employed as in \cite{9652996}, and the integration of both AoU and Data Shapley value based on the two distinct policies outlined in \ref{Designing a Scheduling Policy}. These figures effectively highlight the performance variations among these scheduling approaches. The results indicate that our proposed algorithm significantly enhances the optimization of the FL training process. Notably, the impact of criterion \enquote{AoU or Data Shapley value} is most pronounced in the initial training stages, for instance, when the communication rounds are less than $30$. This observation demonstrates the scheme's ability to expedite the convergence of FL and subsequently enhance learning efficiency.

\begin{figure*}
  \centering
  \subfigure[]{\includegraphics[width=0.32\textwidth]{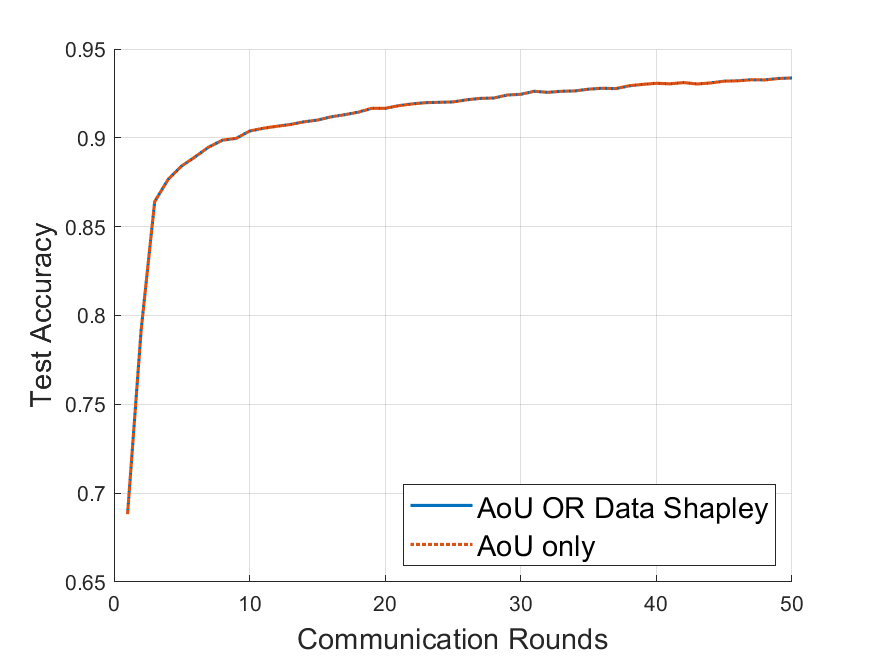}}
  \hfill
  \subfigure[]{\includegraphics[width=0.33\textwidth]{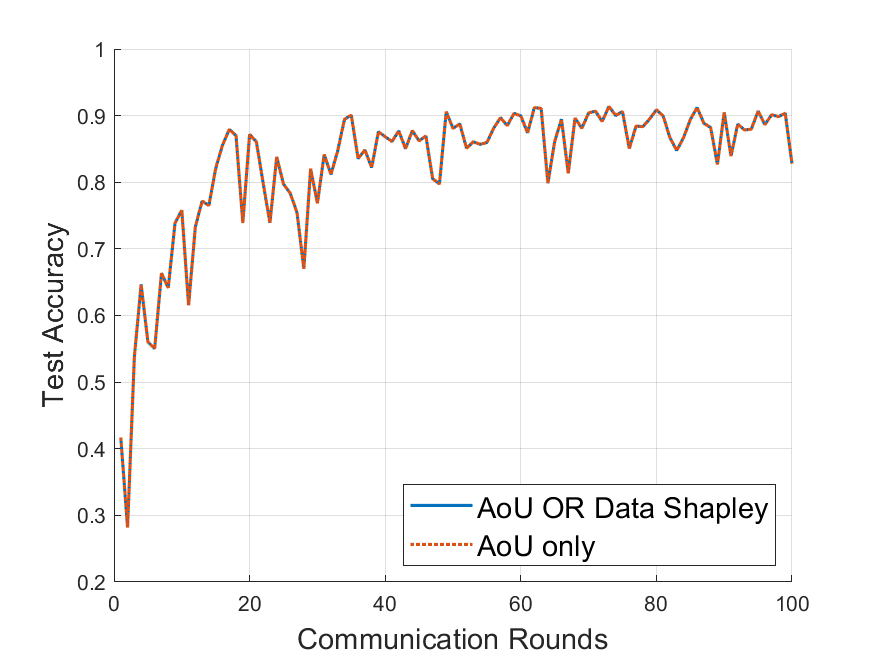}}
  \hfill
  \subfigure[]{\includegraphics[width=0.33\textwidth]{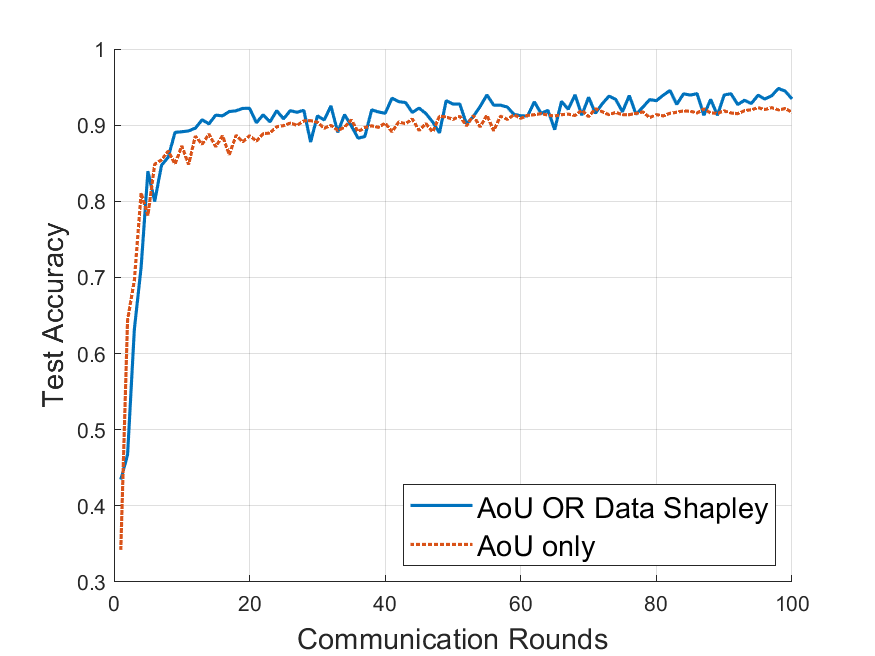}}
  \caption{Comparison of FL training convergence rates under AoU OR Data Shapley vs AoU-only scheduling policies, trained on MLP, using MNIST dataset, $N = 30$, (a) $p=0.1$, i.i.d client assignment (b) $p=0.1$, non-i.i.d client assignment (c) $p=0.8$, non-i.i.d client assignment.}
  \label{fig:figure4}
\end{figure*}

Fig.\ref{fig:figure4}(a, b) emphasizes that in instances of unreliable communication channels (probability of $0.1$), neither the scheduling policy utilized nor the available number of communication channels influences the convergence rate, as indicated by consistent results. Fig.\ref{fig:figure4}c shows the comparison between AoU and AoU-OR-Data Shapley based scheduling in case of non-i.i.d distribution of MNIST data set with reliable channel conditions with $p=0.8$.

\section{Conclusions}
This study delved into the effectiveness of training Federated Learning models over wireless networks. Our approach introduces a novel scheduling policy that optimizes the convergence rate of FL by considering factors like value and freshness of information. We consider important system parameters such as the likelihood of reliable transmission and limitations in spectral resources. Our method employs Age of Update (AoU) and Data Shapley value to gauge update staleness and value. Our study has demonstrated that when clients establish dependable connections with the server in each communication round, accelerating the convergence rate of FL model training becomes achievable by prioritizing fresh and valuable updates. These insights are valuable for improving the performance of FL systems within mobile edge networks.

We aim to enhance scheduling efficiency through adaptive thresholds for Age of Update (AoU) and Data Shapley, which respond to real-time network conditions. Further research can investigate the viability of our solutions on resource-constrained devices with a focus on energy-efficient scheduling. Real-world deployments in sectors like healthcare, finance, and industry will provide valuable insights into the challenges and possibilities of implementing our scheduling solutions.

\bibliographystyle{IEEEtran}
\bibliography{references.bib}
\end{document}